\UseRawInputEncoding
\documentclass[aps,twocolumn,superscriptaddress,showkeys,showpacs,longbibliography]{revtex4-2}
\usepackage{amsmath}
\usepackage{pifont}
\newcommand{\pj}[1]{\textcolor{red}{#1}}
\usepackage{amssymb}
\usepackage{etoolbox}
\usepackage{mwe}
\usepackage{graphicx}% Include figure files
\usepackage{dcolumn}% Align table columns on decimal point
\usepackage{xcolor}
\usepackage{bm}% bold math
\usepackage{makecell,comment}
\usepackage{float}
\usepackage{feynmp}
\usepackage{hyperref}
\usepackage[capitalize]{cleveref}
\hypersetup{
    colorlinks=true,
    linkcolor=blue,
    filecolor=blue,      
    urlcolor=blue,
    citecolor=blue,
    pdftitle={main},
    pdfpagemode=FullScreen,
    }
\usepackage[all]{hypcap}    %for going to the top of an image when a figure reference is clicked
\usepackage{multirow}

\begin{document}

\preprint{APS/123-QED}

\title{A Landau Theory for Pair Density Modulation in Fe(Te,Se)  flakes. }% Force line breaks with \\
%\thanks{A footnote to the article title}%

\author{Po-Jui Chen}
 \affiliation{%
 Department of Physics and Astronomy, Rutgers University, Piscataway, New Jersey 08854, USA 
}%

\author{Piers Coleman}%
 \email{Second.Author@institution.edu}
\affiliation{%
 Department of Physics and Astronomy, Rutgers University, Piscataway, New Jersey 08854, USA
}%
\affiliation{Department of Physics, Royal Holloway University of London, Egham, Surrey TW20 0EX, United Kingdom}%Lines break automatically or can be forced with \\

\date{\today}% It is always \today, today,
             %  but any date may be explicitly specified

\begin{abstract}
Motivated by recent scanning tunneling microscopy (STM) experiments reporting a pair-density modulation (PDM) in flakes of FeTe${_{0.55}}$Se${_{0.45}}$, we develop a Landau theory to elucidate its physical origin. We analyze the PDM in terms of screw and glide symmetries, interpreting it as a hybridized state of two order parameters with opposite glide and screw parity. 
To explain the absence of PDM in the bulk, we argue that the breaking of glide symmetry at the surface allows nematic order to selectively stabilize the PDM in thin flakes. From these symmetry constraints, we show that the opposing glide and screw parities of the condensate favor a site-based, rather than bond-based, pairing mechanism. Thus the discovery of PDM in superconducting flakes suggests that the pairing in iron-based superconductors is local to the iron atoms, possibly driven by Hunds coupling. {\color{black} We argue that the mismatch in Knight shift between the even and odd parity order parameters will lead to a  magnetic-field enhancement of the PDM at low fields, and a reentrant triplet phase at high fields that can be tested in STM experiments.} 
    
\end{abstract}

\maketitle

\begin{comment}
    The understanding of superconductivity is a central challenge in condensed matter physics. In particular, there are several quantum material whose superconducting behavior goes beyond the standard BCS theory. 
\end{comment}

\begin{figure}
\center
\includegraphics[width=1\linewidth]{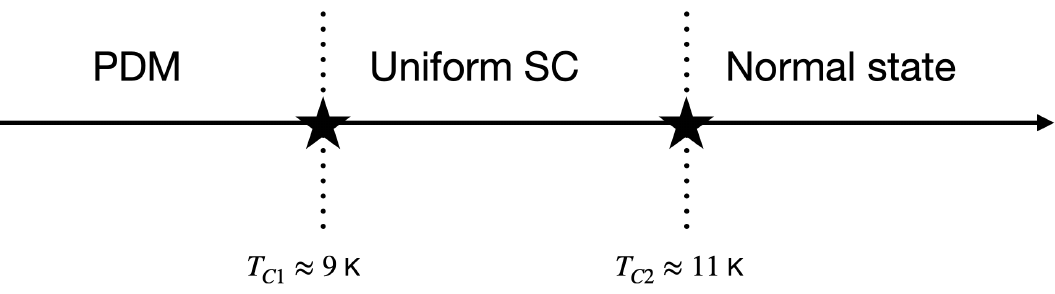}
    \caption{Schematic illustration of the two-step phase transition. The ground state evolves from PDM state into a uniform superconducting order and then into a normal state. The two critical temperatures are $T_{c1}\approx 9$K and $T_{c2}\approx 11$K \cite{kongCooperpairDensityModulation2025} .
}
     \label{PDM_phase}
\end{figure}

\begin{figure}[h]
\center
\includegraphics[width=1\linewidth]{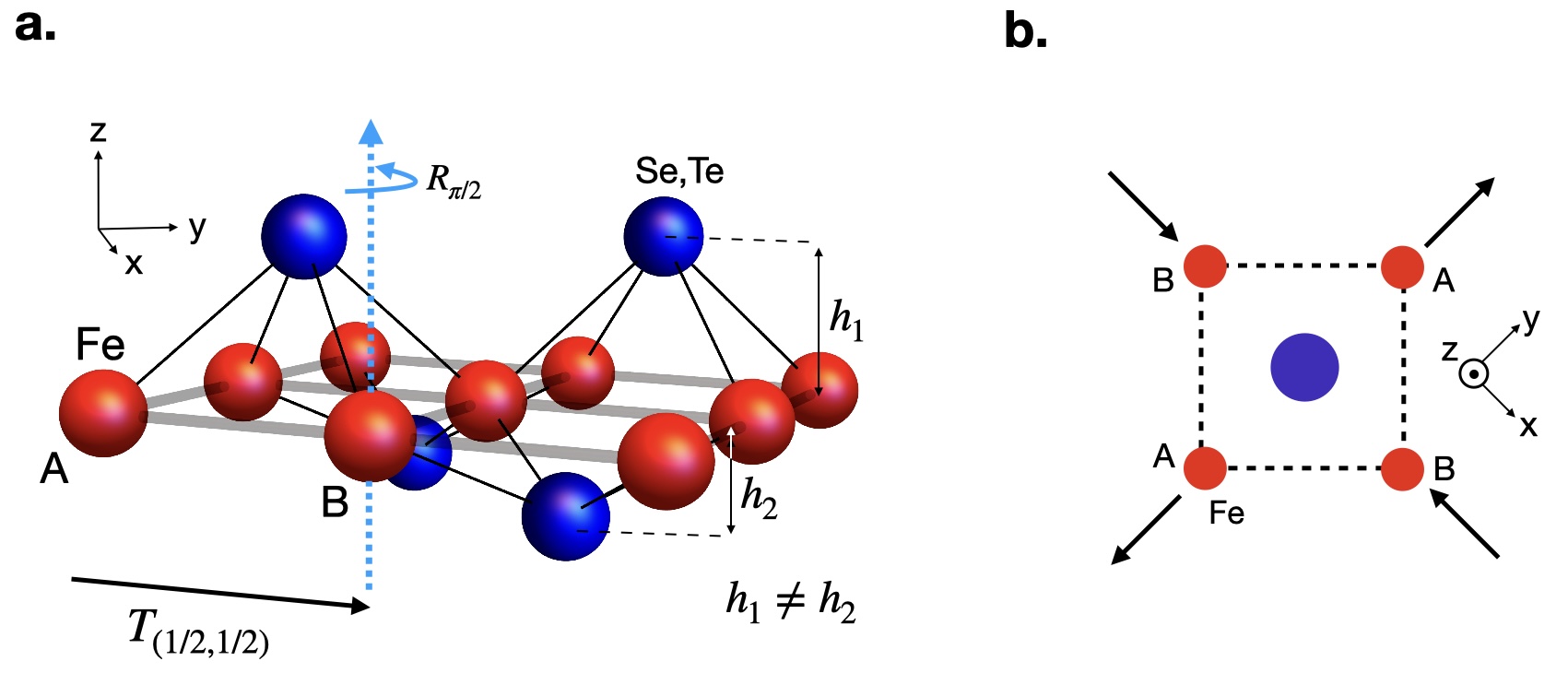}
    \caption{{\bf a.} Structure of single-layer Fe(Te,Se)  showing the different distance of the upper and lower chalcogenide atoms from the iron plane. {\bf b.} atomic displacements associated with nematic order, which  change sign under a $\pi$ rotation.  }
     \label{lattice}
\end{figure}

\textit{Introduction.} 
Iron-based superconductors \cite{coldeaElectronicNematicStates2021,fernandesLowenergyMicroscopicModels2017,hanaguriUnconventionalWaveSuperconductivity2010,chubukovPairingMechanismFeBased2012,mazinUnconventionalSuperconductivitySign2008,stewartSuperconductivityIronCompounds2011} provide a fertile platform for exploring a wide variety of correlated electronic phases. Numerous experiments point towards an unconventional origin for the superconductivity in these materials \cite{mazinUnconventionalSuperconductivitySign2008,liuPairDensityWave2023,fernandesIronPnictidesChalcogenides2022}  \pj{}. 
%Among the many discoveries, one particularly intriguing phenomenon is the pair-density-wave (PDW) state\cite{liuPairDensityWave2023} in which  the superconducting order parameter breaks translational symmetry and develops spatial modulation. %To date, such PDW states are rare, as twisting the phase of the order parameter costs kinetic energy. 
%The typical PDW is induced by a pre-existing broken symmetry in the normal state, such as  a Zeeman split Fermi surface \cite{fuldeSuperconductivityStrongSpinExchange1964,Larkin:1964wok} or charge density-wave order\cite{daiPairdensityWavesChargedensity2018,wangInterplayPairChargedensitywave2015}. 
A new kind of  modulated superconducting state was discovered in recent scanning tunneling microscopy (STM) experiments\cite{zhang2024visualizinguniformlatticescalepair,ding2024sublatticedichotomymonolayerfese,kongCooperpairDensityModulation2025,weiWeiObservationSuperconductingPair2025} on ultra-thin flakes of FeTe${_{0.55}}$Se${_{0.45}}$, which  reveal\cite{kongCooperpairDensityModulation2025} that the superconducting gap differs by up to 40\% on the two iron sublattices, forming a ``{\sl pair density modulated state}'' (PDM). Two sequential superconducting transitions are observed on cooling\cite{kongCooperpairDensityModulation2025}:  first, uniform superconductivity develops at $11$K, then a second transition 
into the PDM state occurs at  $9$K (see Fig.\ref{PDM_phase} (a)). In the PDM, the gap, defined by one half the peak-to-peak separation in the density of states, becomes different at the two iron sites, A and B, in the unit cell.  The gap modulation $(||\Delta_A|-|\Delta_B||)/(|\Delta _A| + |\Delta_B|)= f$ is $20\%$ in the sample principally studied in experiments, but rises to $40\%$ in the thinnest flakes, while  entirely vanishing  in the thickest samples.  Moreover, the modulated gap structure is also accompanied by a nematic order, forming domains in which the  square iron plaquets undergo a rhombohedral compression(see Fig. \ref{lattice}(c)) of about 1\% along their diagonals.

The PDM is intriguing because it resembles a pair-density-wave (PDW) state\cite{liuPairDensityWave2023} in which  the superconducting order parameter also develops spatial modulation. %To date, such PDW states are rare, as twisting the phase of the order parameter costs kinetic energy. 
However, unlike the PDW, the PDM does not break translational symmetry: in this sense it is the superconducting counterpart 
 of the altermagnet\cite{Song2025Altermagnets}, for both preserve lattice translational symmetry, because of an underlying non-symmorphic crystal structure with two equivalent magnetic or superconducting atoms per unit cell.   A second distinct feature of the PDM is that it develops
spontaneously, in the absence of  a pre-existing broken symmetry in the normal state, such as  a Zeeman split Fermi surface \cite{fuldeSuperconductivityStrongSpinExchange1964,Larkin:1964wok} or charge density-wave order\cite{daiPairdensityWavesChargedensity2018,wangInterplayPairChargedensitywave2015} that often accompany PDWs.

%\textit{Symmetry consideration.} 
In bulk FeSe$_{1-x}$Te$_x$ the iron atoms form stacked square lattice layers with the  chalcogenide Se or Te atoms alternately located above or below neighboring iron plaquets. This gives rise to a non-symmorphic structure  described by  space group P4/nmm(No.129)\cite{fischerSuperconductivityLocalNoncentrosymmetricity2011,huIronBasedSuperconductorsOddParity2013} with two iron atoms per unit cell.  The structure is invariant  under a glide ($ G_z$) \footnote{{Technically, this is a glide-mirror, a combination of a half translation and a mirror reflection. For ease of description, we shall  simply adopt the term "glide"}} or a screw operation ($\tilde C_4$): 
\begin{equation}
{G}_z=M_z T_{(1/2,1/2)}, \qquad {\tilde C}_4=R_{\pi/2} T_{(1/2,1/2)}    
\end{equation} 
corresponding to half-lattice translations  $T_{(1/2,1/2)}$   followed by  a reflection in the $xy$ plane $(M_z)$ or a  $\pi/2$ rotation about the z-axis $ (R_{\pi/2})$ (see Fig.\ref{lattice}(a)).  However, in  thin flakes, the chalcogens  lie at different distances above and below  the iron-plane(see Fig. \ref{lattice}), removing the mirror symmetry\cite{kongCooperpairDensityModulation2025}, leaving  just the screw symmetry to protect the equivalence of the two iron atoms. It is this symmetry that is broken by the PDM order. Many experimental observations have reported the existence of nematic order $\Phi$ \cite{ishidaPureNematicQuantum2022,fernandesIronPnictidesChalcogenides2022} in Fe(Te,Se) or monolayer FeSe, which also breaks the screw symmetry which must therefore be considered as an important element in the development of the PDM. 

In a recent paper,  Papaj, Kong, Nadj-Perge and Lee (PKPL)\cite{papaj} propose a BCS description of the PDM state based on the loss of the glide  symmetry.  In their model, the development of a hybrid order parameter $s_{\pm }+ d$ between an extended s- and d-wave superconducting phase  removes the equivalence between the two lattice sites and giving rise to a modulated gap structure.   While the PKPL theory provides an illustration of a PDM, the role of the symmetries are hidden within the structure of the internal tight-binding model and the assumed BCS couplings. In particular,  the screw symmetry of the flakes does not play an explicit role in their theory. 

%Moreover, the connection with lattice nematicity is absent and the close vicinity between the transition temperatures of the $s_{\pm}$  and $d-$wave phases is a finely-tuned feature of the model which needs to be understood.  

Here, we develop a complementary description of the PDM based on Landau theory:  our philosophy is to use the experimental properties of the PDM to gain insight into the microscopic physics.  Landau theory allows us to observe the role of order-parameter symmetry and its interplay with nematic order; in particular, the screw symmetry of the flakes plays a central role. We are particularly interested in understanding why the PDM is  present in thin flake samples, yet absent in the bulk.  Moreover, our approach allows us to consider the constraints on the underlying pairing mechanism that are imposed by the properties of the PDM. 

%We begin by considering the symmetry constraints on the relevant order parameters, using this to construct a Landau free energy framework which allows us to understand the phase diagram and stability of the PDM state.  

On rather general grounds,  we expect that  the order parameters which mix to produce the PDM are  representations of the screw symmetry, with a well-defined eigenvalue under the screw operation. Since the formation of PDM requires screw symmetry broken, it is natural to treat the PDM phase as a hybridization of two order parameters {\color{black}$\Delta_+$ and $\Delta_-$} of opposite screw parity. Generically, two different order parameters of this kind are separated by a first order phase transition. We argue that their coupling to nematicity  only arises at the surface, and  this stabilizes the hybridized PDM in the thin flakes. This places some rather interesting constraints on the pairing that can be further tested by experiment. 

Under the screw rotation $\tilde{C}_4$, the superconducting order parameter generally transforms as
$\tilde{C}_4^{-1}\Delta \tilde{C}_4 = e^{im\pi/2}\Delta$
, where   $m$ is the angular momentum quantum number. In our discussion, we focus on the time-reversal-symmetric cases with $m=0,2$, for which $\tilde{C}_4^{-1}\Delta_{\pm} \tilde{C}_4 = \pm\Delta_{\pm}$. The transformation properties of $\Delta_+$, $\Delta_-$  and $\Phi$ are listed in  Table \ref{symmetry}. 
%{\color{black} connect $\Delta_{AA},\Delta_{BB}$ to $\Delta_-,\Delta_+$}
 \begin{table}[h]
    \centering
    \renewcommand{\arraystretch}{1.2}
    \begin{tabular}{|c|c|c|c|}
        \hline
       Order Parameter & $\Phi$ & $\Delta_+$ & $\Delta_-$ \\ \hline
        $\tilde{C}_4$ eigenvalues & $-1$ & $+1$ & $-1$ \\ \hline
    \end{tabular}
    \caption{Symmetry of the three order parameters under fourfold screw operation $\tilde{C}_4$.}
    \label{symmetry}
\end{table}

{\color{black} We can use symmetry to relate the gap parameters $\Delta_{A,B}$ on sublattice A and B to the order parameters $\Delta_\pm$. To leading order the onsite gap parameter  $\Delta_{A}$  can be expressed as a linear combination $\Delta_{A} = \alpha\Delta_++\beta\Delta_-$. Under a screw rotation, $C_4^{-1}\Delta_{A}C_4=\Delta_{B}$, so  $\Delta_{B} = \alpha\Delta_{+}-\beta \Delta_{-}$ picks up a relative minus sign.  Adjusting the normalization of the order parameters so that  $\Delta_A=\Delta_++\Delta_-$ and $\Delta_B = \Delta_+-\Delta_-$. It follows that  once the two order parameters coexist, screw symmetry is broken and the gap amplitudes develop a modulation between neighboring iron sites.  }

\textit{Landau free energy}. To describe the PDM, we introduce a Landau free energy that is a function of  nematic order $\Phi$ and two superconducting order parameters of opposing parity $\Delta_+,\Delta_-$,
\begin{equation}
F = F_{\Delta}+F_{\Delta\Phi}, 
\end{equation}
where
\begin{equation}\label{FE}
    \begin{split}
        F_{\Delta} &= a [(T-T^+_{c})\Delta_{+}^2+
(T-T^-_{c})\Delta_{-}^2]\cr &+\frac{1}{2}(u_+\Delta_{+}^4+ u_-\Delta_{-}^4+2 u_{\pm}\Delta_{+}^2\Delta_{-}^2),\\
%F_\Phi&= \Phi^2,\\
F_{\Delta\Phi} &= \alpha \Phi^2+ \lambda\Phi\Delta_+\Delta_-.
    \end{split}
\end{equation}
Here  $F_{\Delta}$ is the superconducting (SC) part of the free energy, while $F_{\Delta\Phi}$ describes the nematic order and its coupling to the superconductivity.   $T^+_{c}$ and $T^-_{c}$ are the intrinsic  transition temperatures of the two superconducting order parameters, tuned by an external field $g$, so that $T^{\pm}_{c} = T_0 \pm g $  where $g=0$ defines a multicritical point where these transition temperatures are equal.  $a>0$ and $u_-,u_\pm$ and $u_+$ define the repulsion between the SC order parameters. Quartic terms that contain an odd power of $\Delta_+$ or $\Delta_-$ are not invariant under the screw symmetry and are excluded. Finally note that we assume that the two order parameters are phase-locked, with a single overall phase that can be factored out of the free energy, so that the $\Delta_\pm$ are real (See \footnote{
{\color{black}The possibility of a relative phase $\phi$ between the two condensates requires  introduction of a biquadratic Josephson coupling term $-u_J (\Delta_+\Delta_-)^2\cos(2\phi)$ into the interactions. However, so long 
 as $u_J>0$ is positive, the phases of the order parameters do indeed lock together in the ground-state, with $\phi = 0,\pi$ corresponding to the two screw-degenerate ground-states,  and $u_J$ can be re-absorbed into a redefinition of $\tilde u_\pm=u_\pm - u_J$. (see Supplementary materials for details).}} for more details.)
 
\begin{comment}
    A general expression of the coupling term takes the form $F_{coupling}\sim\Phi^\alpha\Delta_-^\beta\Delta_+^\gamma$. To determine exponents $\alpha,\beta$ and $\gamma$, we perform the symmetry analysis again. Recall that the Symmetry allowed term should respect the symmetry of the normal state where $\Delta_-,\Delta_+$, and $\Phi$ are absent. Thus the coupling term should be invariant under screw rotation symmetry but does not preserve gliding mirror symmetry. Under the gliding mirror reflection, this term picks up the factor $(-1)^\gamma$. To avoid the gliding mirror invariance, one should impose $\gamma$ to be an odd number. On the other hand, the coupling term acquires $(-1)^{\alpha+\gamma}$  under a screw rotation. Ensuring invariance under this operation imposes the constraint that $\alpha+\gamma$  is even. To satisfy the above two conditions, the minimal choice is $\alpha=\beta=\gamma=1$ ane the coupling part of the free energy 
\end{comment}

The second term $F_{\Delta \Phi}$ describes the nematic order and its coupling to the superconductivity. FeTe$_{0.55}$Se$_{0.45}$ lies  in close proximity to a nematic quantum critical point\cite{ishidaPureNematicQuantum2022,zhaoNematicTransitionNanoscale2021}, but  does not spontaneously develop nematic order  outside the PDM, permitting us to approximate the nematic component by a simple quadratic function $\alpha\Phi^2$, where  $\alpha>0$ denotes the distance from the nematic critical point. The minimal symmetry-allowed coupling $\Phi\Delta_+\Delta_-$ is the only term permitted by the screw symmetry, given $\Phi$ changes sign under screw rotations. 
%For instance, combinations such as \Delta_+\Delta_-$ or $\Phi^2\Delta_-\Delta_+$ are  not invariant under a screw rotation. 

\begin{figure}[h]
   \center
    \includegraphics[width=1.0\linewidth]{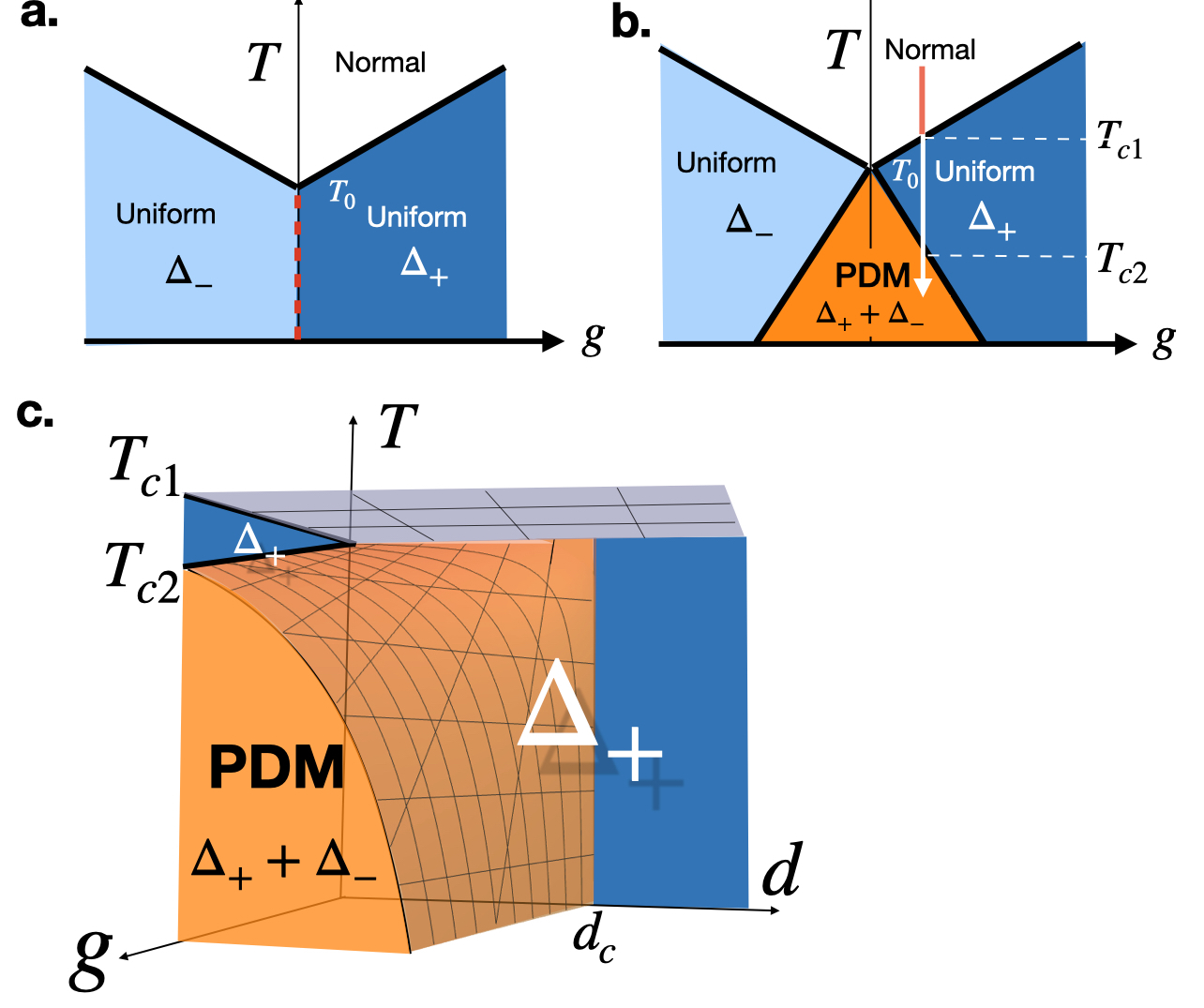}
    \caption{ Phase diagram of the Landau theory showing {\bf a.} when $u_{\pm}\geq U_c = \sqrt{u_+u_-}$, where the co-existence region is absent and {\bf b.} where $u_{\pm}<U_c = \sqrt{u_+u_-}$ and a hybridized phase develops beneath a tetracritical point. {\bf c.} representative 3D phase diagram assuming a nematic coupling to the order parameters, exponentially dependent on flake thickness $d$,  calculated  using  $u_\pm=8,u_+=u_-=2,\lambda_0 = 7,\alpha =1, \xi = 1,T_0=1$. }
     \label{GL_phase}
\end{figure}

Our Landau theory highlights the importance of nematic fluctuations.  Generically, the phase boundary between the  the even and odd-parity phases, depends on the strength $u_{\pm}$ of the repulsion between the order parameters(a detailed derivation is provided in the supplementary materials): if $u_{\pm} > U_c$ exceeds the critical value 
\begin{equation}\label{condition}
   U_c = \sqrt{u_+u_-}.
\end{equation} 
the transition between the two phases is first order as shown in Fig.\ref{GL_phase}{\bf a}, but if $u_{\pm}$ is less than this value, a tetracritical point is formed and a hybridized co-existence region opens up within the phase diagram,  as shown in Fig.\ref{GL_phase}{\bf b}.
Now if we ``integrate out" the nematic degree of freedom, rewriting 
\begin{eqnarray}  \label{shift}
 F_{\Delta\Phi} &\rightarrow &\alpha \biggl( \Phi + \frac{\lambda}{2\alpha}\Delta_+\Delta _-\biggr)^2,\cr
u_{\pm}&\rightarrow &u_{\pm}^* =u_\pm-\frac{\lambda^2}{4\alpha},\end{eqnarray}
we see that nematic fluctuations lower the repulsion between the $\Delta_+$ and $\Delta_-$ degrees of freedom, while also predicting that nematic order will develop as a secondary order parameter in a  PDM with $\Phi = -\frac{\lambda}{2\alpha}\Delta_+\Delta _-$ \cite{szaboSuperconductivityinducedImproperOrders2024}.

This suggests a natural  explanation for the absence of the PDM in bulk FeTe$_{0.55}$Se$_{0.45}$ is that the nematic $\Phi\Delta_-\Delta_+$ is forbidden {by the additional symmetries of the bulk}. In the bulk, both glide-mirror and inversion symmetries are restored and  $\Phi$ is even parity under both operations, so if $\Delta_{\pm}$ have {\sl opposite} parities under either symmetry,  the nematic coupling will vanish in the bulk. Provided that the system is  close enough to the nematic critical point, (small $\alpha$), so that 
\begin{eqnarray}\label{lam_condi}
   \frac{\lambda^2}{4\alpha} >  u_{\pm } -\sqrt{u_+u_-}>0,
\end{eqnarray}
 then by \eqref{shift},  $u_{\pm}> U_c$ in the bulk, but acquires the nematically suppressed value $u^*_{\pm}< U_c$  in the flakes, promoting PDM order. As the flakes get thicker, glide symmetry  will be restored; modeling the  dependence of the  nematic coupling $\lambda$ on thickness $d$ as  $\lambda(d) =\lambda_0 e^{-\frac{d}{\xi}}$, where $\xi$ is the structural healing length, then using \eqref{lam_condi} we can identify an upper-critical flake thickness for the  PDM 
 \begin{eqnarray}
     d_c = \frac{\xi}{2}\ln\biggl[\frac{\lambda^2_0}{4\alpha(u_\pm-\sqrt{u_+u_-})} \biggr].
 \end{eqnarray}
 Fig. \ref{GL_phase}{\bf c.} shows the associated phase diagram. {\color{black} In principle, we could have  added   gradient terms,  $|\nabla \Delta_\pm|^2$ and $(\nabla \Phi)^2$ to the Free energy (Ginzburg Landau theory). Our simplified Landau theory describes the limit in which the superconducting and nematic coherence lengths are short compared with the  healing length $\xi$.
 While gradient terms are needed to describe details of the domain walls and spatial structure, we do not expect them to change the basic phase diagram and symmetry constraints we have derived.  }

\textit{Discussion--} 
Let us comment on the possible superconducting pairing configurations that can give rise to the pair-density modulated state. One  candidate\cite{papaj} is a bond-based pairing, in which Cooper pairs consist of electrons mainly residing at  different lattice sites. Representative bond-based order parameters are extended-s and d-wave \begin{eqnarray}
    \Delta(k) = 
    \left\{\begin{array}{ll}
    \Delta_+(\cos k_x+\cos k_y)     &    \ \hbox{s$_{\pm}$}\cr
       \Delta_-(\cos k_x-\cos k_y)  & \ \hbox{d},
    \end{array}\right.
\end{eqnarray}
where the x-and y-axes lie along the {\sl diagonal} of the iron plaquettes\cite{papaj}. However,  in the bulk, these Cooper pair symmetries transform {\sl identically} under inversion or glide-mirror, so  a nematic coupling  $\lambda \Phi \Delta_-\Delta_+$  is permitted in both flakes and the bulk. {\color{black}Thus  bond-based pairing, though widely used in spin-fluctuation theories of iron-based superconductors\cite{fernandesLowenergyMicroscopicModels2017,mazinUnconventionalSuperconductivitySign2008,fernandesIronPnictidesChalcogenides2022}, does not provide a natural explanation for the marked absence of PDM in  bulk samples (see Table \ref{Mz_sym}).}

An intriguing alternative  is local pairing \cite{andersonFurtherConsequencesSymmetry1985,hazraTripletPairingMechanisms2023,komijaniTripletPairingOrbital2025,colemanTripletResonatingValence2020,Lee18}, in which pairs develop  at the iron sites. {\color{black} Coulomb repulsion between electrons on the iron sites tends to rule out local s-wave pairing, so  the alternative is triplet pairing, likely driven by the strong Hunds coupling inside the iron atoms.}
In this case, a minimal description involves a competition between a uniform (in-phase)  and a staggered (out-of-phase) pairing state\cite{hottaOddParityTripletPair2004,andersonFurtherConsequencesSymmetry1985} within the 2-Fe unit cell. These have opposite glide and screw parities, restricting nematic coupling to the flakes where glide symmetry is absent. Representative order parameters  take the form
\begin{eqnarray}\label{lts}
    \underline{\Delta}(\vec x)= \left\{\begin{array}{ll}\vec{\Delta}_+ \cdot \vec \sigma (\sigma_2\lambda_2)& \hbox{uniform}\cr
  \vec{\Delta}_- \cdot \vec \sigma (\sigma_2\lambda_2)e^{i\vec Q\cdot \vec x}& \hbox{staggered},
    \end{array}\right.
\end{eqnarray}
where $\vec x$ is the spatial co-ordinate of the iron sites, 
 $\vec \Delta_{+,-}$ are the d-vectors for the uniform and staggered triplet pairing, $(\lambda_2)_{\alpha\beta}= -(\lambda_2)_{\beta\alpha} $ is the antisymmetric operator acting on t$_{2g}$ orbital states, e.g.  $\alpha, \beta \in (d_{zx},d_{zy})$, while $\vec Q = (2\pi, 0)\equiv (0,2\pi) $ is the wave-vector for the PDM. { We note that the re-establishment of an inversion center between the Fe-atoms in the bulk ensures that in the presence of spin-orbit coupling,  {\color{black} on the Fermi surface}, the uniform $\Delta_+$ phase is an even parity singlet, while the bulk staggered $\Delta_-$ phase is an odd-parity triplet\cite{andersonFurtherConsequencesSymmetry1985}.}(see Table \ref{Mz_sym}).

\begin{table}[h]
    \begin{tabular}{|c|c|c|c|}
\hline
\multirow{3}{*}{$(\Delta_+,\Delta_-)$}
  & \multirow{3}{*}{$G_z$}& \multirow{3}{*}{$P$}
  & \multirow{3}{*}{\begin{tabular}{c} Bulk $\Phi\Delta_+\Delta_-$ \\ symmetry allowed? \end{tabular}} \\ 
  & & &\\ 
  & & &\\ \hline
($s_\pm$-wave, d-wave) & $(1,1)$ &(+,+)  &   $\checkmark$    \\ \hline
(uniform, staggered)   & $(1,-1)$ &(+,\ -) & $\times$  \\ \hline
\end{tabular}
    \caption{ Bulk glide  ${G_z}$ and spatial parity eigenvalues $P$  for bond-based and local pairing scenarios, showing that a bulk  nematic coupling is forbidden for local  pairing. }
    \label{Mz_sym}
\end{table}

{\color{black} One of the most striking aspects of the PDM implied by Landau theory, is its vicinity to a tetra-critical point. While this might reflect an accidental degeneracy between two unrelated order parameters, the symmetry-based  local pairing perspective  provides a more natural explanation. In this case we would expect that the BCS coupling constants of the two phases to originate from the same local atomic interactions, so the tetracritical point  can be understood as the point where their pair susceptibilties are equal. 
%Indeed, this mechanism for   pair density modulation could also occur in other non-centrosymmetric superconductors  under the tuning effect  of pressure. An interesting candidate for future consideration is UTe$_2$\cite{Aoki2019,Lewin2023,Coleman_Panigrahi_Tsvelik_2022}, which exhibits a  pressure-induced superconducting multi-critical point\cite{ran2020enhancement}. 

We end with a discussion of the predicted effect of magnetic field on the PDM in FeSe$_{1-x}$Te$_x$.  Here, an interesting precedent is set by the heavy fermion compound CeRh$_2$As$_2$,  which exhibits a magnetic field-tuned phase transition from a singlet to a triplet superconductor\cite{khimFieldinducedTransitionSuperconducting2021,landaetaFieldAngleDependenceReveals2022}, thought to involve a change from uniform to staggered superconducting order \cite{landaetaFieldAngleDependenceReveals2022,Lee_Agterberg_Brydon_2025}. If we regard  CeRh$_2$As$_2$ as iron selenide without the nematicity, then we 
can use its example to anticipate the  behavior of thin flakes of FeSe$_{1-x}$Te$_x$ in a magnetic field. 
The suppression $\chi_+$ of the spin susceptibility or Knight shift of the singlet $\Delta_+$ phase is expected to be substantially larger than the suppression $\chi_-$ in the $\Delta_-$ triplet phase. We can then model this physics by adjusting the quadratic terms of the Landau theory\eqref{FE}, writing
\begin{equation}\label{thefield}T_c^{\pm }(B)=T_0 \pm g  -  \chi_{\pm}B^2,\end{equation}where $ \chi_+ \gg \chi_-$. This corresponds to a mapping  
$T_0\rightarrow T_0 - \frac{1}{2}(\chi_++\chi_-)B^2$ and $g\rightarrow g - \frac{1}{2}(\chi_+-\chi_-)B^2$:
in other words, a differential Knight shift will mean the magnetic field acts as the tuning parameter. The corresponding field-temperature phase diagram is then a conformal transformation of Fig. \ref{GL_phase} b which preserves its structure  and tetra-critical point, as  shown in Fig. \ref{figx} (see supplementary materials for details). At large fields, the Pauli-limitation of the singlet $\Delta_+$ phase induces a second-order transition into the uniform triplet $\Delta_-$ phase. Moreover,  since $\Delta_{A,B} = \Delta_+\pm \Delta_-$, the gap modulation parameter $f=(\bigl||\Delta_A|-|\Delta_B|\bigr|)/(|\Delta _A| + |\Delta_B|)= {\rm min}( \frac{\Delta_-}{\Delta_+},\frac{\Delta_+}{\Delta_-})$  is predicted to increase with field (see Fig. \ref{figx}  inset ), peaking in the center of the PDM phase, before returning to zero in the re-entrant triplet phase.  Of course, our treatment gives us no information about the direction of the d-vector of the underlying order, so the relevant direction for applying the field  can not be inferred from the Landau theory.  However, the main expectations are a robust prediction of the local triplet scenario for FeSe$_{1-x}$Te$_x$ \eqref{lts} that can be tested in future STM measurements.  }

\begin{figure}[h]
   \center
    \includegraphics[width=0.9\linewidth]{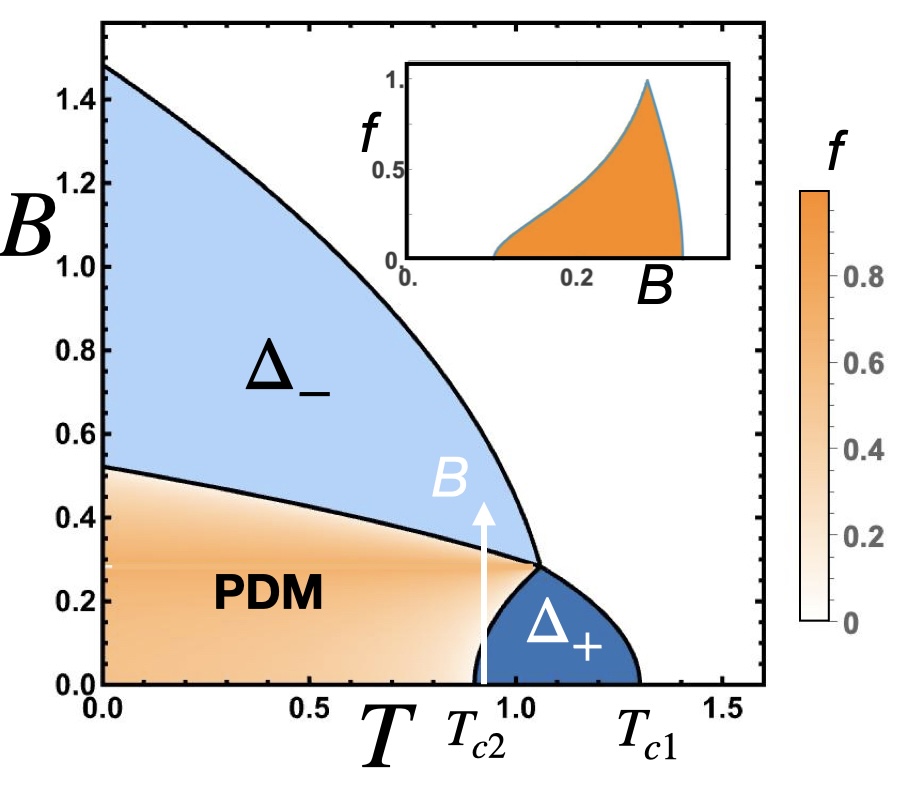}
    \caption{Schematic field-temperature phase diagram predicted from Landau theory, according to \eqref{thefield}, showing field-stabilization of gap-uniform, $\Delta_-$ phase.  Orange density 
    labels the gap modulation parameter $f=\bigl|(|\Delta_A|-|\Delta_B|)\bigr|/(|\Delta _A| + |\Delta_B|)$. Inset: field dependence of $f$ in PDM for field labeled by white arrow. Parameters used were $u_+=u_-=2,u_{\pm}=1,T_0=1.2, g=0.1,T_{c1}=1.3,T_{c2}=0.9,\chi_+=3;\chi_-=0.5$. }
     \label{figx}
\end{figure}

\begin{acknowledgments}
    The authors would like to thank Phil Brydon, Po-Yao Chang, Indra Gankuyag, Andreas Gleis, Daniel Kaplan, Liam L.H. Lau, Aaditya Panigrahi, and Kaustubh Roy for discussions related to this project. This work was supported by the Office of Basic Energy Sciences, Material
		Sciences and Engineering Division, U.S. Department of Energy (DOE)
		under Contract DE-FG02-99ER45790.
\end{acknowledgments}

%\bibliographystyle{apsrev4-2}
%\bibliography{FeSC_REF}
%apsrev4-2.bst 2019-01-14 (MD) hand-edited version of apsrev4-1.bst
%Control: key (0)
%Control: author (8) initials jnrlst
%Control: editor formatted (1) identically to author
%Control: production of article title (0) allowed
%Control: page (0) single
%Control: year (1) truncated
%Control: production of eprint (0) enabled
%

\onecolumngrid
\newpage
\makeatletter 

\begin{center}
    \textbf{\large Supplementary materials}\\[10pt]
\end{center}

\setcounter{figure}{0}
\setcounter{section}{0}
\setcounter{equation}{0}

\appendix

\section{Complex formulation of the Landau theory and the reduction to real form}

The complexified Landau theory 
$F = F_{\Delta}+F_{\Delta\Phi}$, is given by 
\begin{equation}
    \begin{split}
        F_{\Delta} &= a [(T-T^+_{c})|\Delta_{+}|^2+
(T-T^-_{c})|\Delta_{-}^2|]\cr &+\frac{1}{2}\biggl(u_+|\Delta_{+}|^4+ u_-|\Delta_{-}|^4+2 u_{\pm}|\Delta_{+}|^2|\Delta_{-}|^2
-u_J(\Delta^{*2}_+\Delta^{2}_-+{\rm H.c}) \biggr),\\
%F_\Phi&= \Phi^2,\\
F_{\Delta\Phi} &= \alpha \Phi^2+ (\lambda/2)\Phi(\Delta^*_+\Delta_- + {\rm H.c}).
    \end{split}
\end{equation}
The complex version of the free energy must allows for a Josephson quartic term of strength $-u_J$ which couples $\Delta_+^2$ and $\Delta_-^2$ in the interactions. Generically, we  expect $u_J>0$, because a Josephson coupling  minimizes the inter-condensate kinetic energy, and microscopically has a minus sign resulting from  virtual quasiparticle excitations.   A quadratic coupling $\Delta_+^*\Delta_- + {\rm H.c}$ is forbidden, except in combination with the nematicity $\Phi$. By writing $\Delta_+ \rightarrow \Delta_+ e^{i\theta}$, $\Delta _+ \rightarrow \Delta_- e^{i(\theta+\phi)}$ where $\Delta_\pm$ are now real and $\phi$ is the relative phase, the free energy can be written in a more compact form
\begin{equation}
    \begin{split}
        F_{\Delta} &= a [(T-T^+_{c})\Delta_{+}^2+
(T-T^-_{c})\Delta_{-}^2]\cr &+\frac{1}{2}\biggl(u_+\Delta_{+}^4+ u_-\Delta_{-}^4+2 \Delta^{2}_+\Delta^{2}_- (u_{\pm}-u_J \cos 2 \phi)\biggr),\\
%F_\Phi&= \Phi^2,\\
F_{\Delta\Phi} &= \alpha \Phi^2+ \lambda\Phi\Delta_+\Delta_- \cos \phi .
    \end{split}
\end{equation}
Minimizing the energy with respect to the nematic order parameter $\Phi$, the resulting free energy takes the form 
\begin{eqnarray}
    F &=& a [(T-T^+_{c})\Delta_{+}^2+
(T-T^-_{c})\Delta_{-}^2]\cr &+&\frac{1}{2}\biggl(u_+\Delta_{+}^4+ u_-\Delta_{-}^4+2 \Delta^{2}_+\Delta^{2}_- \bigl(u_{\pm}+u_J- (\frac{\lambda^2}{4\alpha}+ 2u_J)\cos^2\phi\bigr)\biggr),
\end{eqnarray}
 Minimizing the free energy with respect to the relative phase angle $\phi$ between the two order parameters reveals that the stable state is $\phi = 0,\pi$ provided  $2u_J+\lambda^2/(4\alpha)>0$, so provided $u_J$ is positive, a time-reversal violating state is not expected and the effect of the Josephson coupling can be entirely absorbed by reverting to the real-formulation of the Free energy \eqref{FE} in the main text, with a renormalized value of $\tilde u_\pm = u_\pm - u_J$.

\section{Landau theory: Derivation of Phase boundaries}
\begin{figure}[h]
   \center
    \includegraphics[width=1.0\linewidth]{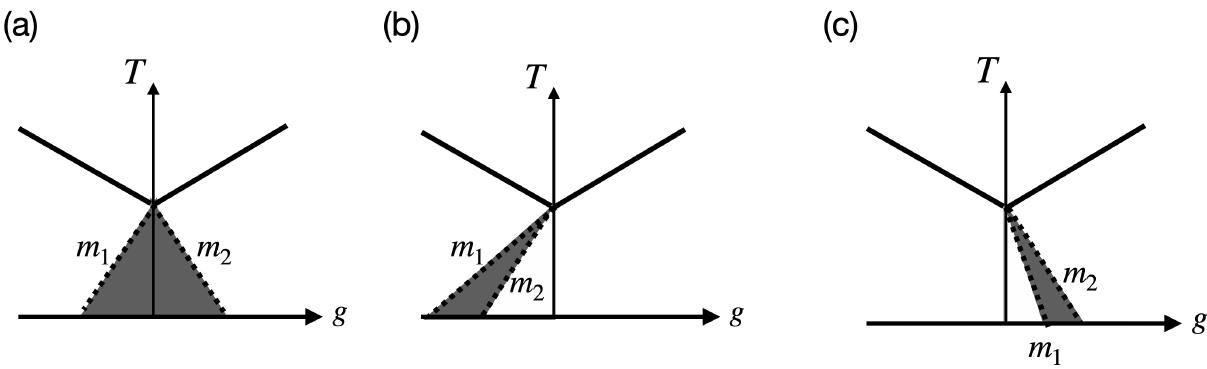}
    \caption{Three possible scenarios for the parity-mixing region in $g-T$ space. (a)$u_+, u_- >u_\pm,$ , (b)$u_+<u_\pm<u_-$ and (c) $u_+>u_\pm >u_-$}
     \label{SOL}
\end{figure}

We recall the effective Landau free energy, which only includes superconducting orders
\begin{equation}
\begin{split}
     F[\Delta_{+},\Delta_{-}] = 
    & a (T-T^+_{c})\Delta_{+}^2+a 
(T-T^-_{c})\Delta_{-}^2\\
&+\frac{1}{2}(u_+\Delta^4_++2u_{\pm}\Delta^2_-\Delta^2_++u_-\Delta_-^4)
\end{split}
\end{equation}
Using a change of the variable $T^\pm_{c}=T_0\pm g$
. In the co-existence regime, where both $\Delta_+$ and $\Delta _-$ are finite, minimization of the free energy leads to the following two equations
\begin{equation}\label{zero}
    \begin{split}
       (a(T-T_0-g)+u_+\Delta_{+}^2+u_\pm\Delta_{-}^2)=0\\
        (a(T-T_{0}+g)+u_-\Delta_{-}^2+u_\pm\Delta_{+}^2)=0\\
    \end{split}
\end{equation}
Solving \eqref{zero}, we obtain
\begin{eqnarray}
    \Delta^2_+ &=& a\left(\frac{u_--u_{\pm}}{u_+u_--u_\pm^2}\right)\biggl[T_0-T+\left(\frac{u_\pm+u_-}{u_--u_\pm}\right)g\biggr]\\
    \Delta^2_- &=& a\left(\frac{u_+-u_{\pm}}{u_+u_--u_\pm^2}\right)\biggl[T_0-T-\left(\frac{u_\pm+u_+}{u_+-u_\pm}\right)g\biggr]
\end{eqnarray}
The condition $\Delta_+\rightarrow 0$  determines the phase boundary 1 into the $\Delta_-$ phase on the left-hand side of the phase diagram below.  Likewise, the condition $\Delta_-\rightarrow 0$  determines the phase boundary into the $\Delta_+$ phase on the right-hand side of the phase diagram.  This then sets two phase boundaries 
\begin{eqnarray}
T_0-T^{(L)}_{c2}+\left(\frac{u_\pm+u_-}{u_--u_\pm}\right)g=0 & \Rightarrow & T^{(L)}_{c2}= T_0+\left(\frac{u_\pm+u_-}{u_--u_\pm}\right)g\qquad \hbox{(Phase boundary 1)}\cr
T_0-T^{(R)}_{c2}-\left(\frac{u_\pm+u_+}{u_+-u_\pm}\right)g=0&\Rightarrow&T^{(R)}_{c2}=T_0-\left(\frac{u_\pm+u_+}{u_+-u_\pm}\right)g\qquad \hbox{(Phase boundary 2)}
\end{eqnarray}
As $u_{\pm}$ is increased, the left-hand and right-hand second order phase boundaries converge into a single line, and for larger values of $u_{\pm}$ the co-existence phase is replaced by a single first-order phase boundary.  There are three scenarios, as illustrated in Fig. \ref{SOL} below

The two phase boundaries merge when the gradients become equal. If $u_+\neq u_-$ , then as $u_{\pm} $ is increased, it must exceed min$(u_-,u_+)$, leading to scenario (b) or (c), so that the criteria for the two phase boundaries to merge is then
\begin{equation}
   m_1= \left(\frac{u_\pm+u_-}{u_--u_\pm}\right)= m_2 = -\left(\frac{u_\pm+u_+}{u_+-u_\pm}\right).
\end{equation}
which implies \begin{equation} u_{\pm}^2= U_c^2 = u_+u_-.\end{equation}   
In the special case where $u_+ =u_-$, the criterion for merging of the two phases is the same. At the point of merger, when $u_{\pm} = \sqrt{u_+u_-}$, the joint gradient of the phase boundary is then
\begin{equation}\label{grad}
    m_1=m_2 = m= \frac{dT_c}{dg}=\frac{\sqrt{u_+}+\sqrt{u_-}}{\sqrt{u_-}-\sqrt{u_+}}.
\end{equation}

For larger values of $u_{\pm}> U_c$, the equilibrium values of the single order parameters are given by 
\begin{equation}
    \Delta_{\sigma} = \sqrt{\frac{a(T_c^{(\sigma)}-T)}{u_{\sigma}}},\qquad \Delta_{-\sigma }= 0,  \qquad \sigma = (-,+),
\end{equation}
and the corresponding free energy is then given by 
\begin{equation}
    F_{\sigma}= - a^2\biggl[\frac{(T_0-T+\sigma g)^2}{2u_{\sigma}}\biggr],\qquad \sigma = (-,+).
\end{equation}
The first order phase boundary separating the two phases occurs when $F_+ = F_-$, which then determines the phase boundary given by  
\begin{equation}
\frac{(T_0-T_c)+g}{\sqrt{u_+}}=\frac{(T_0-T_c)-g}{\sqrt{u_-}},
\end{equation}
or 
\begin{equation}
(T_c-T_0) = \left(\frac{\sqrt{u_+}+\sqrt{u_-}}{\sqrt{u_-}-\sqrt{u_+}}\right)g,
\end{equation}
from  which we see that the gradient $m= dT_c/dg$ of the first order boundary matches the gradient found at the point of fusion between the two second-order phase boundaries \eqref{grad}. 

\section{Field-temperature phase diagram }

\begin{figure}[h]
   \center
    \includegraphics[width=0.35\linewidth]{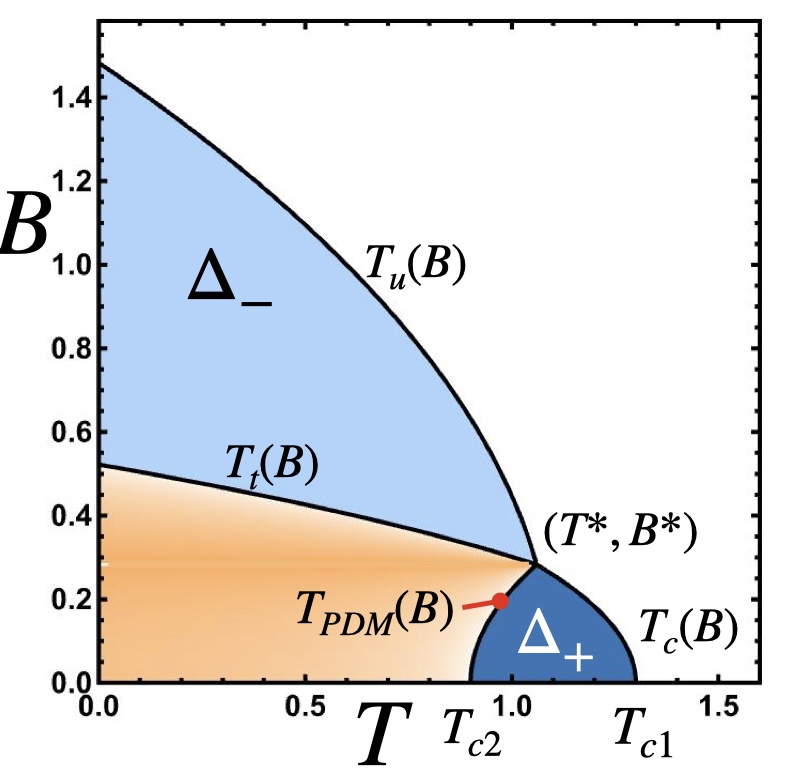}
    \caption{Schematic showing the four transition temperatures \eqref{transt} predicted in a field 
    using the Landau  Theory. Parameters used were $u_+=u_-=2,u_{\pm}=1,T_0=1.2, g=0.1,T_{c1}=1.3,T_{c2}=0.9,\chi_+=3;\chi_-=0.5$. }
     \label{figfin}
\end{figure}

As discussed in the main text, the addition of a magnetic field is modelled by the field-dependent correction of the transition temperatures
\begin{equation}
T_c^{\pm} = T_0\pm g-\chi_{\pm}B^2, 
\end{equation}
where  $\chi_{\pm}$ determine the strength of the Knight-shift for the $\pm$ phases. We assume $\chi_+\gg\chi_-$ if the $\Delta_-$ is a triplet order parameter. If we write $\chi_{\pm} = \chi \pm \Delta \chi$, 
with $\chi = \frac{1}{2}(\chi_++\chi_-)$ and $\Delta \chi = \frac{1}{2}(\chi_+-\chi_-)$
the this corresponds to the replacements
\begin{eqnarray}
    T_0&\rightarrow& T_0- \chi B^2, \cr
    g&\rightarrow& g- \Delta \chi B^2.
    \end{eqnarray}
and with these substitutions, we can generalize all the results of Appendix B to finite $B$. 
For convenience, we now list the main results for the special case where $u_+=u_-=u$,  
$T_{c1}=T_0+g,\quad
T_{c2}=T_0-\alpha g$,  $\alpha = (u+u_{\pm})/(u-u_{\pm})$  and
\begin{eqnarray}\label{transt}
T_{c}(B)&=&T_0 + g - \chi_+B^2\cr
T_{PDM}(B)&=&T_0 -\alpha g + ( \alpha \Delta \chi -\chi)B^2\cr
T_{t}(B)&=&T_0 +\alpha g - ( \alpha \Delta \chi +\chi)B^2\cr
T_{u}(B)&=&T_0 - g - \chi_-B^2
\end{eqnarray}
where $T_c(B)$ is the Pauli-limited transition temperature of the uniform superconductor, $T_{PDW}(B)$ is the field-dependent PDM transition temperature, $T_t(B)$ is the re-entrant temperature temperature for the pure triplet phase ($\Delta_-$) and $T_u(B)$ is the Pauli-limiting  field of the $\Delta_-$ phase  as illustrated in Fig. \ref{figfin}.
The location of the tetracritical point is at $(T^*,B^*)$, where
\begin{equation}
    T^* = T_0 - g \left(\frac{\chi}{\Delta \chi}\right), \qquad B^* = \sqrt{\frac{g}{\Delta \chi}}.
\end{equation}

\end{document}